\documentclass{desyproc}

\begin{document}
\title{Search for exotics at $BABAR$}

\author{{\slshape Elisabetta Prencipe$^1$ on behalf of the $BABAR$ Collaboration}\\[1ex]
$^1$Forschungszentrum J\"ulich, Leo Brandt Strasse, 52428 J\"ulich, Germany \\(previously addressed at JGU, University of Mainz, Germany)}
\contribID{xyz}

\confID{8648}  
\desyproc{DESY-PROC-2014-04}
\acronym{PANIC14} 
\doi  

\maketitle

\begin{abstract}
One of the most intriguing puzzles in hadron spectroscopy are 
the numerous charmonium-like states observed in the last decade, 
including charged states that are manifestly exotic. The $BABAR$  experiment
has extensively studied those in B meson decays, initial state radiation 
processes and two photon reactions.  The study of the process 
$B \rightarrow J/\psi \phi K$,  with a search for the 
resonant states X(4140) and X(4270) in their decays to $J/\psi \phi$, 
will be highlighted. The recent results of the Dalitz analysis 
of $\eta_c$ to 3 pseudoscalar mesons, via 2-photon interactions,  
will be presented in this report [Contribution talk: ID 201].
 
\end{abstract}

\section{Introduction}
Several new Charmonium-like states have been observed at $BABAR$, revealing a spectrum too rich to be uniquely described by potential models\cite{NRPM}. Different hypotheses have been proposed from theorists to explain their nature,  such as hybrid charmonium states, diquark-antidiquark states or $D^0 \bar D^{0(*)}$ molecules\cite{modellicharmonio}.  
 The QCD spectrum is much richer than that of the naive quark model, as the gluons, which mediate the strong force between quarks, can also act as principal components of entirely new types of hadrons. These gluonic hadrons fall into two general categories: glueballs (excited states of pure glue) and hybrids (composed by a quark, an antiquark, and excited glue).  The additional degrees of freedom carried by gluons allow glueballs and hybrids to have spin-exotic quantum numbers $J^{PC}$ that are forbidden for normal mesons and other fermion-antifermion systems. Exotic quantum numbers (e.g. $0^{--}$, $0^{+-}$, $1^{-+}$, $2^{+-}$) are the easiest way to distinguish gluonic hadrons from $q \bar q$ states. 
Predictions for hybrids come mainly from calculations based on the bag model, flux tube model, and constituent gluon model and recently, with increasing precision, from Lattice QCD. New forms of matter, such as glueballs or molecular states, are predicted by QCD to populate the low mass region of the hadron mass spectrum\cite{antimo3}. This  motivates the study of  the $J/\psi$ radiative and hadronic decays.  

Two analyses will be shortly summarized in this report: the Dalitz analysis of $\eta_{c} \rightarrow K^+ K^- \eta$/$\pi^0$ via 2-photon interactions and the study of the invariant mass systems of $J/\psi \phi$,  $J/\psi K$ and $KKK$ in $B$ decays. The first is relevant to several issues in light meson spectroscopy, and it is recently published by $BABAR$\cite{antimopapaer}. No Dalitz plot analysis has been performed on $\eta_C$ three-body decays until now. The second presents a new determination of the branching fraction (BF) of $B^{\pm,0} \rightarrow J/\psi K^+ K^- K^{\pm,0}$ and $B^{\pm,0} \rightarrow J/\psi \phi K^{\pm,0}$, using eight times more data than that reported by the PDG\cite{BBarArticle}, and search for exotic states.

\section{Decay of $\eta_c \rightarrow K^+K^- \eta$/$\pi^0$ via 2-photon interactions}
Recently, a search for exotic resonances was performed by $BABAR$ through Dalitz plot analyses of $\chi_{c1}$ states\cite{antimo4}. Scalar mesons are still a puzzle in light-meson spectroscopy, as there are too many states and they are not consistent with the quark model. In particular, the $f_0(1500)$ resonance, discovered in $\bar p p$ annihilations, has been interpreted as a scalar glueball\cite{antimo5}. However, no evidence for the $f_0(1500)$ state has been found in charmonium decays. Another glueball candidate is the $f_0$(1710) discovered in radiative $J/\psi$ decays. Recently, $f_0(1500)$ and $f_0(1710)$ signals have been incorporated in a Dalitz plot analysis of $B \rightarrow 3K$ decays\cite{antimo6}. Charmless $B \rightarrow XK$ could enanche gluonium production\cite{antimo7}. Another puzzling state is the $K^*_0(1430)$, never observed as clear peak in $K \pi$ invariant mass. Its parameters were measured from the LASS experiment in $K^- p \rightarrow K^- \pi^+ n$\cite{antimo8}.

\begin{figure}[htb]
\centering
\mbox{
{\scalebox{0.15}{\includegraphics{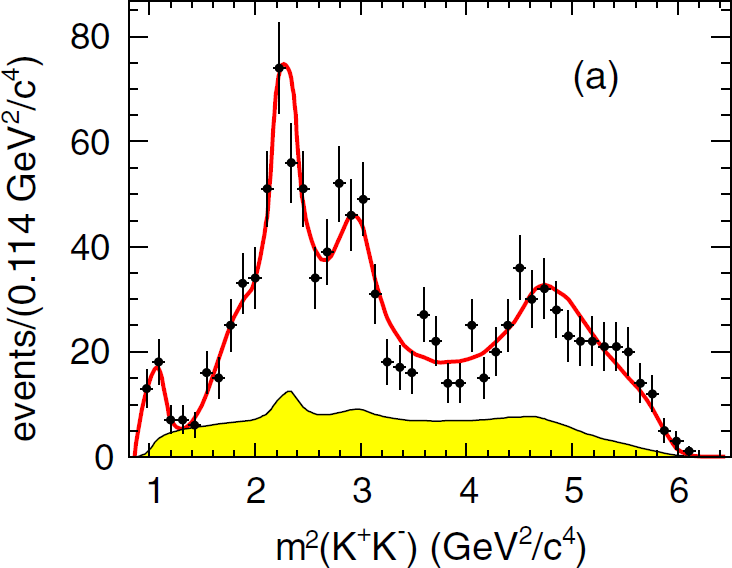}}} \quad
{\scalebox{0.15}{\includegraphics{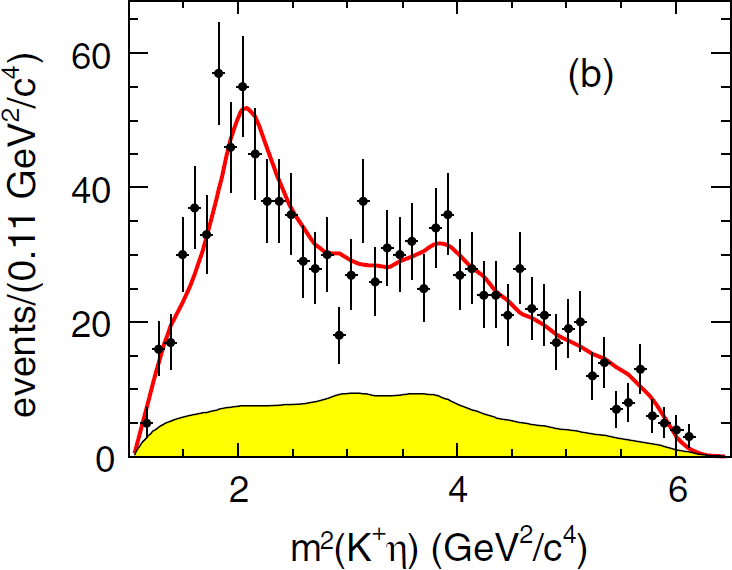}}} \quad
{\scalebox{0.15}{\includegraphics{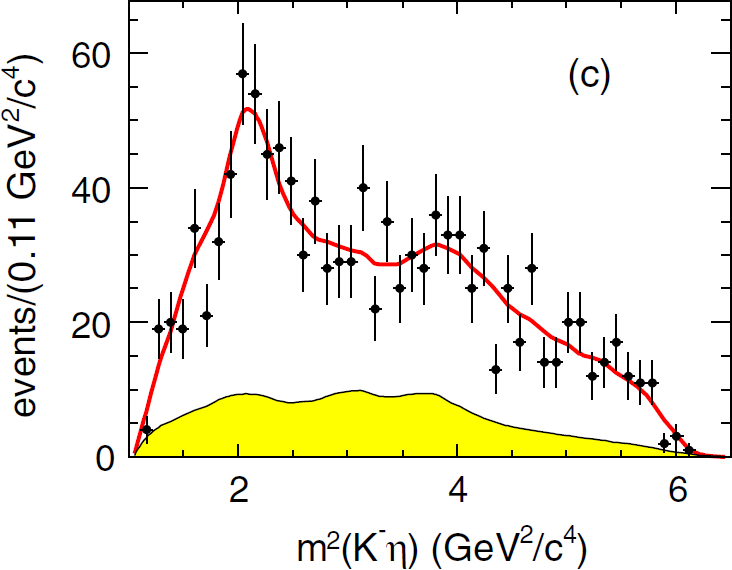}}}
}
\caption{\label{Fig10-babar} Projections from the  Dalitz plot of $\eta_c \rightarrow K^+K^- \eta$. The shaded (yellow) histograms show the estimated background. The state $K^*_0(1430)$ is seen as clear peak in (b) and (c).}
\end{figure}

\begin{figure}[htb]
\centering
\mbox{
{\scalebox{0.15}{\includegraphics{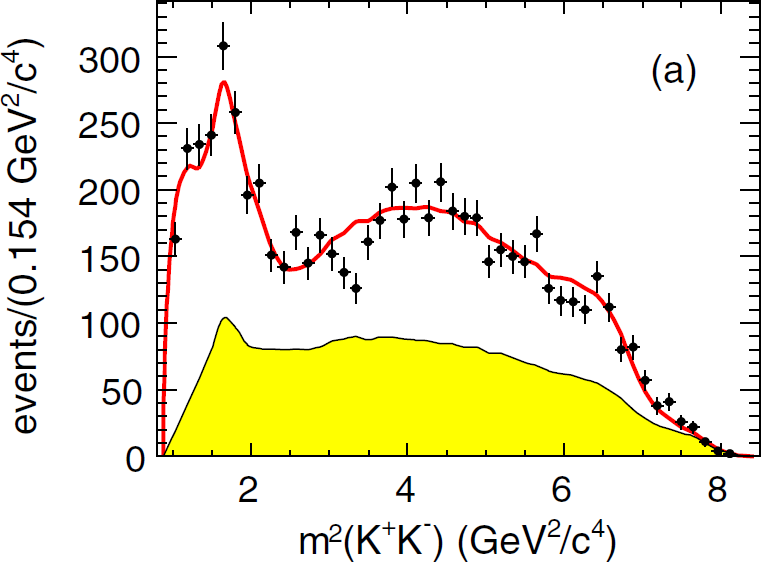}}} \quad
{\scalebox{0.15}{\includegraphics{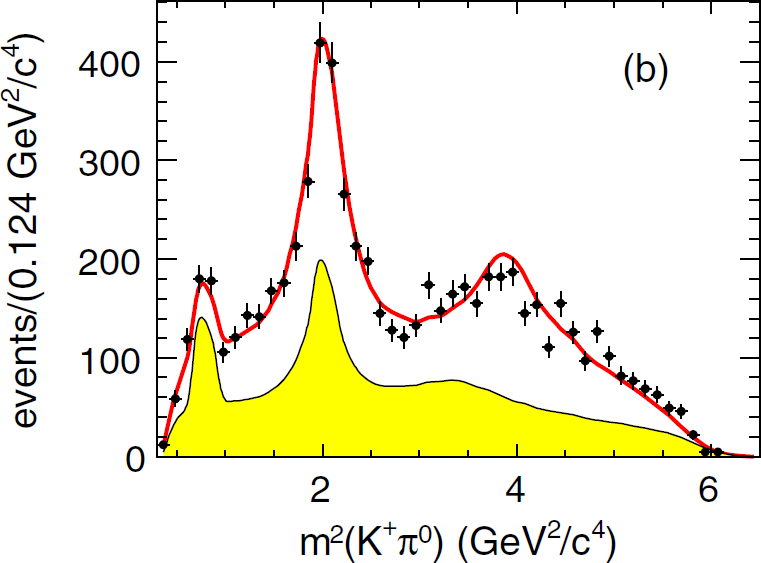}}}  \quad
{\scalebox{0.15}{\includegraphics{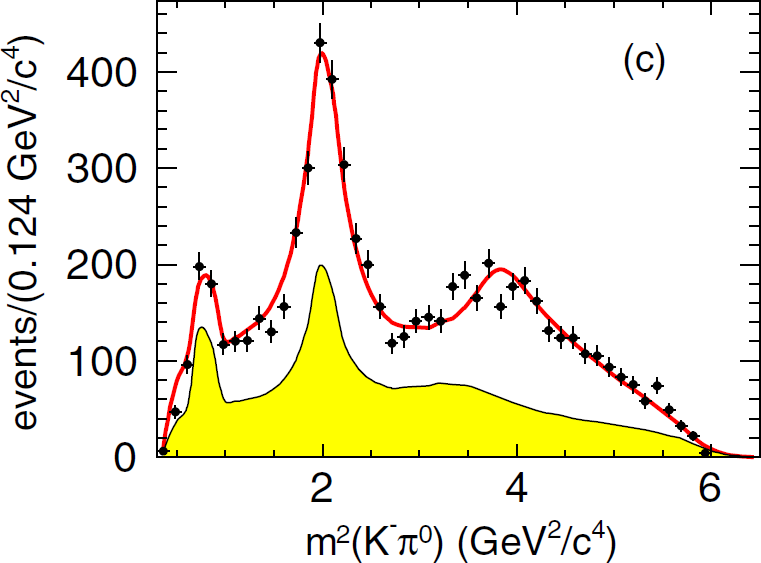}}} 
}
\caption{\label{Fig11-babar} Projections from the Dalitz plot of $\eta_c \rightarrow K^+K^- \pi^0$ . The shaded (yellow) histograms show the estimated background. The state $K^*_0(1430)$ is seen as clear peak in (b) and (c).}
\end{figure}

We describe  a study of the decays $\eta_c \rightarrow K^+ K^- \eta$ and $\eta_c \rightarrow K^+ K^- \pi^0$, with $\eta \rightarrow \pi^+ \pi^- \pi^0$, $\eta \rightarrow \gamma \gamma$ and $\pi^0 \rightarrow \gamma \gamma$, produced in two-photon interactions. The data sample used is 519 fb$^{-1}$ at $BABAR$. Two-photon events in which at least one of the interacting photons is not quasireal are strongly suppressed by a dedicated selection. A clear peak of $\eta_c$ is seen, and well reconstructed in the invariant mass systems of $K^+ K^- \eta$ and $K^+ K^- \pi^0$. The Dalitz analysis is then performed for $\eta_c \rightarrow K^+ K^- \eta$ and  $\eta_c \rightarrow K^+ K^- \pi^0$: the projection of the invariant mass distributions and their unbinned maximum likelihood fit are shown in Fig.~\ref{Fig10-babar} and~\ref{Fig11-babar}. A clear peak at the mass of $K^*_0(1430)$ is observed in both cases, together with other expected structures.  Amplitude parameterization is performed in a standard way for a pseudoscalar meson decaying to 3 pseudscalar mesons. Full interference is allowed among the amplitudes of all resonances in the Dalitz. No evidence for interferences between signal and background is found, so a sum of inchoerent resonances is used for fitting the sidebands. The non-resonant contribution is included in the fit. From our fit, we learn that the model provides an adeguate description of data for $\eta_c \rightarrow K^+ K^- \eta$, while the isobar model does not describe properly the data for   $\eta_c \rightarrow K^+ K^- \pi^0$. Scanning the likelihood as function of the $K^*_0(1430)$ mass and width, we obtain: m($K^*_0(1430)$) = 1438 $\pm$ 8 $\pm$ 4 MeV/c$^2$ and $\Gamma$($K^*_0(1430)$) = 210 $\pm$ 20 $\pm$ 12 MeV.

In this work also the pseudoscalar meson mixing angle is evaluated: $\theta_P$ = (3.1$\rm ^{+3.1}_{-5.0}$)$^o$, and it differs 2.9$\sigma$ deviation from expectations. This issue involves in theoretical discussions where the siglet and octet mixing angle should be considered separately.

\section{Analysis of the decay $B \rightarrow J/\psi K K K$}
Several resonant structures, whose masses are above the $D \bar D$ threshold, are not predicted by potential models. For example, the X(3872) have been seen in $B \rightarrow X K, X \rightarrow J/\psi~ \pi^+ \pi^-$, or Y(4260) was observed by investigating the process $e^+e^- \rightarrow \gamma_{ISR}X$, $X \rightarrow J/\psi \pi^+ \pi^-$\cite{X3872, X3872choi,Y4260babar}; but no indication of new states has been observed in the  $J/\psi~ K^+ K^-$ invariant mass system, until the paper quoted in Ref.~\cite{kai} highlighted the possibility of a couple of resonant states, decaying to $J/\psi\phi$, with $\phi \rightarrow K^+ K^-$ and $J/\psi \rightarrow \mu^+ \mu^-$. These observations are controversial. $Strangeness$ in charmonium seems a sector still to be exploited.  The rare decay $B \rightarrow J/ \psi K K K$, in particular $B \rightarrow J/ \psi \phi K$, is interesting because it is a promising place to search for new resonances, as it proceeds, at quark level, via the weak transition $b \rightarrow c \bar c s$. It could be a quasi 2-body decay, $B \rightarrow X_g K$, with $X_g \rightarrow J/ \psi \phi$, where $X_g$ = $\lvert g c \bar c s \bar s>$, with gluonic contribution ($g$). 

An unbinned  maximum likelihood fit is performed  to extract the yield and calculate the BFs. Detailed explanation on these calculations are presented in Ref.~\cite{elisabetta}, together with the relevant discussion for the non-resonant $K^+K^-$ contribution to the BF of $B \rightarrow J/\psi KKK$ and systematic uncertainty calculation. Here we report only the relevant information for the analysis of the three invariant mass distributions: $J/\psi \phi$, $J/\psi K$, $KKK$, for both $B^+$ and $B^0$ samples. In this analysis we calculate also: 
$R_\phi$ = $\cal B$($B^0 \rightarrow J/\psi \phi K^0_S$)/$\cal B$($B^+ \rightarrow J/\psi \phi K^+$) = 0.48 $\pm$ 0.09 $\pm$ 0.02 , and $R_{2K}$ = $\cal B$($B^0 \rightarrow J/\psi K^+ K^- K^0_S$)/$\cal B$($B^+ \rightarrow J/\psi K^+ K^- K^+$) = 0.52 $\pm$ 0.09 $\pm$ 0.03; we find values in agreement with the expectation of the spectator quark model (e.g., ratio R$\sim$0.5). These are first measurements. For the first time the non-resonant $K^+K^-$ contribution to the BF of $B \rightarrow J/\psi K K  K$ is observed. No evidence of signal is found for $B^0 \rightarrow J/\psi \phi$, in agreement with theoretical predictions: we evaluate UL$<$1.01 $\cdot$ 10$^{-6}$ at 90$\%$ confidence level (CL).
We search for the resonant states reported by the CDF Collaboration in the
$J/\psi \phi$ mass spectrum. The masses and the widths in our fit are fixed to values according to Ref.~\cite{kai}. 
\begin{figure}[htb]
\centering
\mbox{
{\scalebox{0.20}{\includegraphics{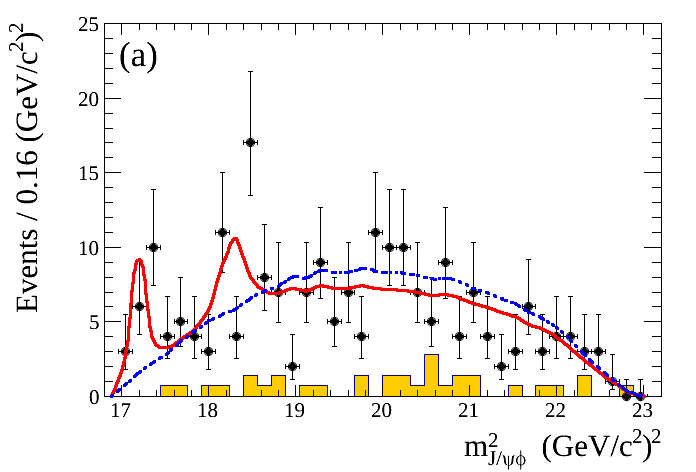}}} \quad
{\scalebox{0.20}{\includegraphics{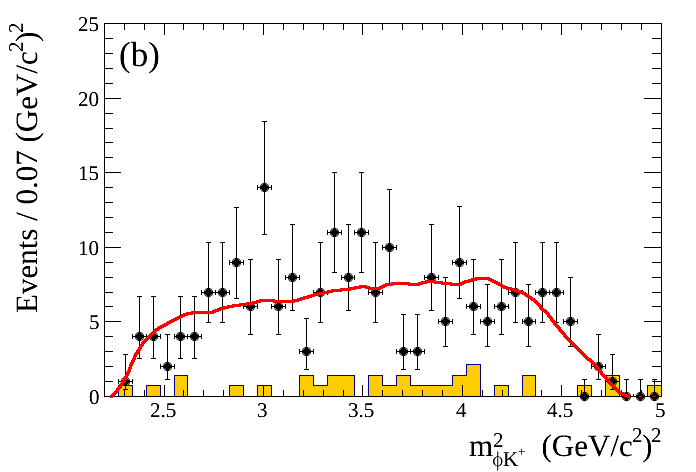}}} \quad
{\scalebox{0.20}{\includegraphics{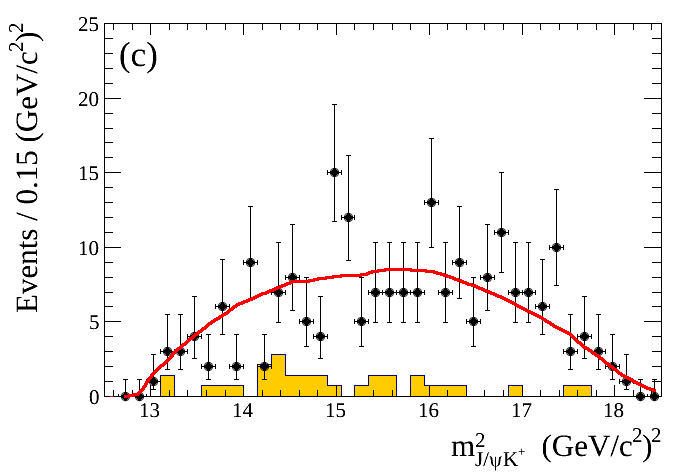}}}
}
\label{Fig20-babar}
\caption{Dalitz plot projections for $B^+ \rightarrow J/\psi \phi K^+$ on (a) $m^2_{J/\psi \phi}$, (b) $m^2_{\phi K^+}$, and (c) $m^2_{J/\psi K^+}$. The continuous (red) curves are the results from fit model performed including the $X(4140)$ and $X(4270)$ resonances. The dashed (blue) curve in (a)  indicates the projection for fit model  with no resonances included in the fit. The shaded (yellow) histograms indicate the evaluated background. Within systematic effects, a significance $<$2$\sigma$ is found for both peaks in (a).}
\end{figure}
 We observed significant efficiency decrease at low $J/\psi \phi$ mass, due to the inability to reconstruct slow kaons in the laboratory frame, as a result of energy loss in the beampipe and SVT material. We model the resonances with an inchoerent sum of two S-wave relativistic Breit-Wigner (BW) functions with parameters fixed to the CDF values~\cite{kai}. A non-resonant contribution is described according to PHSP.  The decay of a pseudoscalar meson to two vector states contains high spin contributions which could generate non-uniform angular distributions. However, due to the limited data sample (212 yield for $B^+$ and 50 for $B^0$, in the signal area, respectively) we do not include such angular terms, and assume that the resonances decay isotropically. The fit function is  weighted by the inverse of the two-dimentional efficiency computed on the Dalitz plots (see the continuous red curve in Fig.~\ref{Fig20-babar}.  Using the Feldman-Cousins method\cite{FC}, we obtain the ULs at 90\% CL: $BF(B^+ \rightarrow X(4140)K^+)\times BF(X(4140) \rightarrow J/\psi \phi)/BF(B^+ \rightarrow J/\psi \phi K^+) < 0.135$ ,  and $BF(B^+ \rightarrow X(4270)K^+)\times BF(X(4270)\rightarrow J/\psi \phi)/BF(B^+ \rightarrow J/\psi \phi K^+) < 0.184$. The $X(4140)$ limit may be compared with the CDF measurement of $0.149\pm 0.039\pm 0.024$~\cite{kai} and the LHCb limit of 0.07~\cite{LHCb}. The $X(4270)$ limit may be compared with the LHCb limit of 0.08. We find that the hypothesis that the events are distributed uniformly on the Dalitz plot gives a poorer description of the data.



\begin{footnotesize}

\end{footnotesize}

\end{document}